\documentstyle[12pt,psfig]{article}

\begin{document}

\begin{titlepage}

\begin{flushright}
MIT-CTP-2384\\
TUTP-94-22\\
gr-qc/9412040\\
November 1994
\end{flushright}

\vskip 0.9truecm

\begin{center}
{\large {\bf A Note on the Semi-Classical Approximation\\
in Quantum Gravity$^*$}}
\end{center}

\vskip 0.8cm

\begin{center}
{Gilad Lifschytz, Samir D. Mathur}
\vskip 0.2cm
{\it Center for Theoretical Physics,
Laboratory for Nuclear Science\\
and Department of Physics,\\
Massachusetts Institute of Technology,
Cambridge MA 02139, USA.\\
e-mail: gil1 or mathur @mitlns.mit.edu }
\vskip 0.3cm
{and}
\vskip 0.3cm
{Miguel Ortiz}
\vskip 0.2cm
{\it Institute of Cosmology, Department of Physics and Astronomy,\\
Tufts University, Medford, MA 02155, USA.\\
e-mail: mortiz@tufts.edu }
\end{center}

\vskip 0.7cm

\noindent {\small {\bf Abstract:} We re-examine the semiclassical
approximation for quantum gravity in the canonical formulation, focusing on
the definition of a quasiclassical state for the gravitational field. It is
shown that a state with classical correlations must be a superposition of
states of the form $e^{iS}$. In terms of a reduced phase space formalism,
this type of state can be expressed as a coherent superposition of
eigenstates of operators that commute with the constraints and so
correspond to constants of the motion. Contact is made with the usual
semiclassical approximation by showing that a superposition of this kind
can be approximated by a WKB state with an appropriately localised
prefactor. A qualitative analysis is given of the effects of geometry
fluctuations, and the possibility of a breakdown of the semiclassical
approximation due to interference between neighbouring classical
trajectories is discussed. It is shown that a breakdown in the
semiclassical approximation can be a coordinate dependent phenomenon, as
has been argued to be the case close to a black hole horizon.  }

\rm
\noindent
\vskip 2 cm

\begin{flushleft}
$^*$ {\small This work was supported in part by funds provided by the
U.S. Department of Energy (D.O.E.) under cooperative agreement
DE-FC02-94ER40818, and in part by the National Science Foundation.}
\end{flushleft}

\end{titlepage}

\section{Introduction}

Although we have few clues about the final form of a quantum theory of
gravity, there is at least one constraint that we can be sure of: That the
theory should, in some limit, describe quantum matter fields interacting
with an essentially classical background spacetime in what is known as the
semiclassical limit of quantum gravity. In some formulations of quantum
gravity, such as perturbation theory around a fixed background, the
semiclassical limit is guaranteed. However, at a more fundamental level we
expect the notion of spacetime to be a derived concept. In any theory
reflecting this feature, it is interesting to understand how the notion of
a classical background emerges.

Some considerable work towards understanding the semiclassical limit has
been done in the canonical approach to quantum gravity using the ADM
formulation \cite{ADM}. In this approach there is no background spacetime,
since the dynamical variables are the 3-metrics of spacelike hypersurfaces,
plus the matter fields on these hypersurfaces.  Employing the Dirac
procedure, physical states must be annihilated by the momentum and
Hamiltonian constraints.  The momentum constraints reduce the configuration
space to the space of all 3-geometries (the equivalence classes
of 3-metrics under spatial diffeomorphisms). The Hamiltonian constraint is
imposed by the Wheeler--DeWitt equation
\begin{eqnarray}
{\cal H}\Psi[h,\phi]=0
\end{eqnarray}
which has the effect of factoring out translations in the time direction,
and is a direct analogue of the Klein-Gordon equation for the quantized
relativistic particle.

The semiclassical limit is obtained by expanding the Wheeler--DeWitt
equation in powers of the gravitational coupling constant $G$.  This really
involves two separate approximations: A WKB approximation (effectively in
$\hbar$) in the gravitational (or more generally classical) sector, and a
Born-Oppenheimer approximation separating the gravity and matter (classical
and quantum) parts. At each order, one performs first the gravitational WKB
approximation, and then solves the remaining equation for the matter
state. In this way the complete solution of the Wheeler--DeWitt equation is
split into a product of a purely gravitational piece representing the
classical geometry and a mixed piece representing quantum field theory on
that background (for simplicity we shall assume that the gravitational
field is the only classical variable).

To first order, the expansion of the Wheeler--DeWitt equation yields the
first WKB approximation
\begin{eqnarray}
	\Psi[h,\phi] = e^{iS_{HJ}[h_{ij}]/\hbar G},\qquad
	{1\over 2}G_{ijkl}{\delta S_{HJ}
	\over \delta h_{ij}}{\delta S_{HJ}\over \delta
	h_{kl}} -2\sqrt{h}R=0
\end{eqnarray}
to the gravitational Wheeler--DeWitt equation, where $S_{HJ}[h_{ij}]$ is a
Hamilton--Jacobi functional for general relativity. We shall refer to a
state of this form, without a prefactor, as a {\it first order WKB
state}. This first order approximation depends only on the
gravitational degrees of freedom.

Any functional $S_{HJ}[h_{ij}]$ that solves the Hamilton--Jacobi equation
defines a congruence of classical trajectories in superspace, made up of all
possible foliations of all solutions of Einstein's equations defined by
$S_{HJ}$ (see Fig. 1).

As was shown by Lapchinski and Rubakov, and later by Banks
\cite{lap,banks}, the next order approximation is obtained by solving two
equations: An equation for the gravitational WKB prefactor $D[h]$, and a
functional differential equation for the remainder of the state $\chi[h,f]$
that depends on the matter variables $f$. Solving the equation for
$\chi[h,f]$ is equivalent to solving the functional Schr\"odinger equation
along each of the congruence of eikonal trajectories corresponding to the
set of all foliations of all solutions of Einstein's equations given by
$S_{HJ}[h_{ij}]$.

\begin{figure}
\centerline{
\psfig{figure=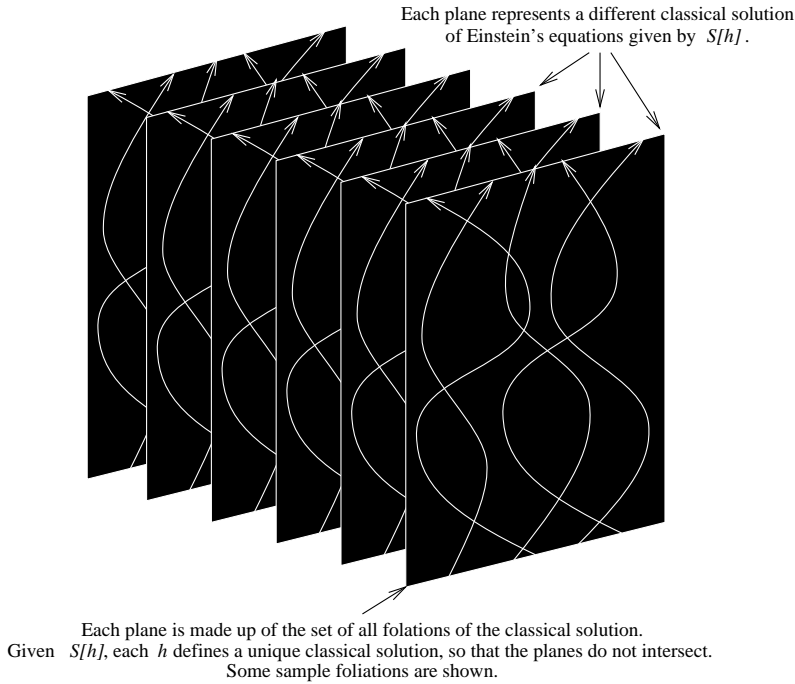,angle=-90,height=9cm}}
\begin{center}
\parbox{5in}{
\caption{\em A congruence of classical solutions in superspace. Each
classical solution is shown as a plane in the diagram, which contains the
different trajectories in superspace corresponding to that solution, given
by different choices of the lapse and shift functions.}}
\end{center}
\end{figure}

The semiclassical approximation to second order should yield a state that
represents a quantum matter field propagating on a quasiclassical
background spacetime. The results we have described are not quite the whole
story, since they do not in themselves provide a complete description of
such a quantum state. A careful treatment of the gravitational degrees of
freedom is needed in order to restrict the semiclassical approximation to
the propagation of quantum matter fields on a {\it single} background.

A semiclassical state with a gravitational WKB part that defines a
congruence of classical trajectories on superspace is not suitable for
describing classical behaviour. In Ref. \cite{jjh}, an approximate form of
the Wigner function was used to argue that a generic first order
gravitational WKB state has classical correlations between coordinates and
momenta. However, it was later pointed out in a simple example that the
exact Wigner function of a general WKB state does not exhibit such
correlations \cite{arley,laflamme}, unless a particular form of the WKB
prefactor is chosen, corresponding to a coherent superposition of first
order WKB states. These results are analogous to the simple statement in
quantum mechanics that a plane wave does not exhibit classical
correlations, but that plane wave states may be superposed to construct a
coherent state.

\subsection{An Outline}

We shall argue in this paper that in quantum gravity, a quasiclassical
state\footnote{We shall use `quasiclassical' to describe a state that
approximates classical behaviour, and `semiclassical' to describe a state
that is a product of a part that is quasiclassical and a part that is not.}
should be a `coherent state' in an appropriately defined sense: The
uncertainty of a complete set of spacetime diffeomorphism invariant
operators (constants of the motion) should be small (a similar proposal was
made some years ago Komar \cite{komar}). We shall show that this condition,
although somewhat formal, can be realised in a concrete way. Since the
classical counterparts of the diffeomorphism invariants together specify a
unique classical solution, this point of view reinforces the notion that a
classical state in quantum gravity gives rise to only an {\it approximate}
background spacetime.  We shall go on to argue that a coherent state can
and must be used as the gravitational state providing a background for the
semiclassical approximation. The approximate nature of the background
described by such a state will be shown to determine some limits of the
semiclassical approximation.

The Hamilton--Jacobi formalism can be used to define a natural complete set
of diffeomorphism invariant operators. Through the Hamilton--Jacobi
formalism, we shall show that a first order WKB state approximates an
eigenstate with respect to half of these operators (the $\alpha$'s of
Hamilton--Jacobi theory). Given this simple observation it follows that a
coherent or quasiclassical state is approximated by a coherent
superposition of first order WKB states.  The rewriting of our
quasiclassical states in quantum gravity as coherent superpositions of
first order WKB states makes direct contact with the work of Gerlach
\cite{gerlach}. He observed that a superposition of WKB states can be
arranged to have support only in a narrow `tube' in superspace around
3-geometries comprising a solution of Einstein's equations. This
observation relates the metric-dependent and coordinate invariant notions
of what constitutes a quasiclassical state. Gerlach's interpretation seems
to have fallen out of favour in recent years, but we hope to revive it in
this paper.

The fact that a coherent state can be written in the WKB approximation
using the metric representation is important for a second reason. It allows
us to use the coherent state in the standard semiclassical approximation as
described above.  An important consequence of the connection between first
order WKB states and $\alpha$ eigenstates is that the expectation value of
conjugate operators (the $\beta$ operators of Hamilton--Jacobi theory) in
such a state is maximally uncertain. Thus a first order WKB state cannot
be expected to have a Wigner function with strong correlations
between coordinates and momenta, and so is not appropriate for describing a
background spacetime. However, a coherent state {\it can} be used as a
background for the semiclassical approximation. It exhibits quasiclassical
correlations and its WKB approximation is precisely a second order WKB
state of the form
\[
\Psi[h_{ij}]={1\over D[h_{ij}]}\, e^{i S_{HJ}[h_{ij}]}
\]
with an appropriately chosen prefactor $D[h_{ij}]$. $D[h_{ij}]$ is given by
rewriting the coherent superposition of first order WKB states as a second
order WKB state. Using the language of Vilenkin \cite{vilenkin}, the WKB
prefactor can be regarded as providing a measure on the space of all
classical solutions compatible with $S_{HJ}[h_{ij}]$. A coherent
superposition is simply a WKB state with this measure concentrated around a
minimally narrow band of classical solutions which effectively represent a
single classical spacetime (the width of such a band is determined by the
condition that the prefactor be of second order in the WKB approximation,
and thus not vary too rapidly)\footnote{One word of warning should be
added: The description of a quasiclassical state as a second order WKB
state gives special prominence to a particular solution $S_{HJ}$ of the
Hamilton--Jacobi equations, and so hides a natural symmetry present in a
coherent superposition. It appears according to this description that
$S_{HJ}$ restricts us to a congruence of spacetimes fixed by specifying
half of the gauge invariant degrees of freedom, and that the prefactor then
restricts the support of the wavefunctional to a small subset of these. The
diffeomorphism invariant description of the state is somewhat more
symmetric, and the wavefunctional is better thought of as having support on
a tube in superspace. There are many classical solutions that run within
this tube, but all have constants-of-the-motion within a narrow spread. No
particular solution (or set of solutions) is preferred as a background for
the semiclassical approximation, but the narrow width of the tube means
that in most circumstances semiclassical physics is not sensitive to the
particular choice of background.}.

After establishing the validity of the semiclassical approximation using
a coherent quasiclassical state, we shall go on to discuss in the last part
of the paper the situations that can lead to a breakdown of this
approximation scheme. Normally, the breakdown of the semiclassical
approximation is thought to be governed by the behaviour of matter fields
on a fixed classical background, with large quantum fluctuations in the
matter fields leading to a loss of classical behaviour. Even in more
careful studies of corrections to the semiclassical approximation
\cite{kiefer}, the emphasis has been on corrections to the equation
describing evolution along individual eikonal trajectories (the functional
Schr\"odinger equation), rather than on interference between neighbouring
trajectories.

The notion that a state in quantum gravity can at best describe a small
neighbourhood of classical backgrounds indicates a simple limitation of the
semiclassical approximation. Recall that the fact that macroscopically
different eikonal trajectories in superspace (corresponding to different
classical solutions) can lead to different matter evolution, underlies the
notion of decoherence between different classical histories. This is
commonly believed to provide a promising explanation of the emergence of
our classical world.  However, because any quasiclassical state describing
the gravitational background has a finite spread, the mechanism responsible
for decoherence can actually spoil classical behaviour \cite{samir-paz}: If
neighbouring geometries contained within the spread of a quasiclassical
state lead to substantially different matter evolution, then the notion of
quantum field theory on a fixed background breaks down. We shall discuss
this aspect of the semiclassical approximation in the last section of this
paper. It is of particular interest since it has recently been shown that
problems of this kind spoil the semiclassical approximation close to a
black hole horizon \cite{samir,klmo,verlindes}.

It is appropriate at this point to mention two caveats to our work. First
of all, we shall not discuss the mechanisms by which one may arrive at the
quasiclassical states described in this paper. Our emphasis is simply on
describing states that could reasonably be called classical and should
therefore arise naturally from initial conditions, decoherence or some
similar mechanism explaining the emergence of classical configurations. For
this reason, we shall also not discuss the possibility of superpositions or
ensembles of quasiclassical states. We refer the reader to the extensive
literature on quantum cosmology for an open ended discussion of some of
these issues. Secondly, the approach we take compares and relates
quantisation in the metric representation to that using the constants of
the motion as the fundamental observables. One would hope that at some
fundamental level these descriptions are equivalent. However, it should be
pointed out that this equivalence is not yet understood. For example, It
has been noted recently \cite{cjz} that there can be more than one way of
defining the Hilbert space in quantum gravity. More precisely, even though
one knows the algebra of operators of a theory, the choice of what states
are physical and what states are unphysical can be made in more than one
way, thus giving inequivalent quantum theories. The resulting quantum
theories are also likely to depend on the classical variables used as a
starting point for quantisation. Nevertheless, a valuable tool in resolving
ambiguities of this kind is to study the semiclassical limit of quantum
gravity, since this gives definite information about the physical spectrum
of the desired theory of quantum gravity. Although it is known that there
are difficulties in operator ordering the Wheeler--DeWitt operator that
block the construction of wavefunctionals \cite{woodard,jackiw}, any
reasonable quantisation in a metric (or connection) representation should
be equivalent to quantisation in terms of constants of the motion, at least
in the semiclassical limit.

\subsection{A simple example}

Before beginning a discussion of WKB states in quantum gravity, it is
useful to run through the definition of a quasiclassical state for the
simple case of the relativistic particle.  For the sake of extreme
simplicity, we shall work with a massless particle in 1+1 dimensional flat
spacetime. The unconstrained phase space for the particle consists of the
spacetime coordinates $x^\mu$ and their conjugate momenta. Time
translations are generated by the constraint $p^2=0$ so that physical
states in the coordinate representation satisfy the Klein--Gordon equation
\begin{eqnarray}
	\hbar^2\partial^\mu\partial_\mu\psi(x^\mu)=0
\end{eqnarray}
In this simple case, we know that the space of physical states is just the
set of all functions $\psi^+(x^+)+\psi^-(x^-)$. However, let us try to find
solutions in the WKB approximation. We write
\begin{equation}
\psi(x^\mu)=e^{iS_0(x^\mu)/\hbar + iS_1(x^\mu) +i \hbar S_2(x^\mu)+\ldots}
\end{equation}
The first term in an expansion in powers of $\hbar$ yields the
Hamilton--Jacobi equation
\begin{eqnarray}
	\hbar^2(\partial_\mu S_0)(\partial^\mu S_0)=0
\end{eqnarray}
for $S_0(x^\mu)$, so that
\begin{eqnarray}
	\psi_0(x^\mu;P,\pm)=e^{iS_0(x^\mu;P,\pm)/\hbar}=
	e^{iP_\mu x^\mu/\hbar}=
	e^{iPx^\pm/\hbar}
\end{eqnarray}
is a first order WKB state. A solution of the Hamilton--Jacobi equation is
given by any function of the form $S_0=P_\mu x^\mu=Px^\pm$, with
$P_\mu=(P,\pm P)$ where $P$ and the choice of sign enter as simple
constants of integration (we are ignoring an additive constant).  $S_0$
then defines the congruence of trajectories with momentum $P_\mu$, since
the Hamilton--Jacobi function assigns a momentum
\begin{equation}
p_\mu={\partial S(x^\mu;P;\pm)\over\partial x^\mu}=P_\mu
\label{mom}
\end{equation}
at each spacetime point. For any value of $P$, $P_\mu=(P,P)$ defines the
congruence of left moving trajectories with momentum $P$ and $P_\mu=(P,-P)$
the right-moving trajectories. By specifying a second constant
\begin{equation}
	X={\partial S_0(x^\mu;P,\pm)\over\partial P}=x^\pm
\label{cons}
\end{equation}
a particular trajectory is selected from the congruence (essentially by
specifying the value of $x$ at $t=0$).  Thus specifying a value for $P$ and
$X$ completely specifies a classical solution. We may invert the relations
(\ref{mom}) and (\ref{cons}) to obtain $X$ and $P$ as functions of $x^\mu$
and $p_\mu$. $X(x^\mu,p_\mu)$ and $P(x^\mu,p_\mu)$ are then constants of
the motion. They thus have weakly vanishing Poisson brackets with the
constraint $p^2=0$ \footnote{Note that for left-moving solutions
$\{X(x^\mu,p_\mu),p^\mu p_\mu\} = \{x^+,p_0^2-p_1^2\} =2(p_0-p_1)$ which
should be regarded as being proportional to the Hamiltonian constraint for
the left-moving sector. Similar reasoning applies to the right-moving
sector.}.  They are also conjugate operators with quantum commutators
$\{X,P\}=1$.

The second order WKB state is obtained by solving the Klein--Gordon
equation to order $\hbar$. From this it follows that
\[
\partial_\mu S_1 \partial^\mu S_0=0
\]
so that $\partial_{\mp}S_1=0$. The imaginary part of $S_1$ becomes the WKB
prefactor so that the second order WKB state is of the form
\begin{equation}
\psi(x^\mu)={1\over D(x^\mu)}e^{iS_0(x^\mu)/\hbar}
	={1\over D(x^\pm)}e^{iPx^\pm/\hbar}
\label{wkb2}
\end{equation}
where $D(x^\mu)=e^{-iS_1(x^\mu)}$, ignoring any lower order correction
to $S_0$.

In order to get a state that is localised in both momentum and position, we
simply construct the relativistic analogue of a coherent state in terms of
the constants of the motion $X(x^\mu,p_\mu)$ and $P(x^\mu,p_\mu)$. These
constants of the motion are precisely the functions of $x^\mu$ and $p_\mu$
that define classical behaviour. In order to define a coherent state, we
must first introduce some scales into the problem. We can study a system
with a given characteristic momentum and characteristic spacetime
resolution whose product is much larger than $\hbar$. We can then define a
Gaussian superposition of first order WKB states of the form
\begin{equation}
\psi(x^\mu) = {1\over(16\pi^3{\bar{P}}^2/\hbar\omega)^{1/4}}
	\int dP e^{-(P-{\bar{P}})^2/2\hbar\omega}
	e^{iP_\mu x^\mu/\hbar}e^{-i(P-{\bar{P}}){\bar{X}}/\hbar}
\label{raw}
\end{equation}
where the last term is a phase that fixes $X$ to be localised around
${\bar{X}}$, and $\omega$ indicates that the uncertainties in position and
momentum need not be equal. There is only constructive interference between
the oscillating functions when (\ref{cons}) is satisfied.

Performing the integral in (\ref{raw}) gives
\begin{equation}
\psi(x^\mu) = \left({\hbar\omega\over 4\pi {\bar{P}}^2}\right)^{1/4}
	e^{-\omega(x^0\pm x^1-{\bar{X}})^2/2\hbar}
	e^{i{\bar{P}}_\mu x^\mu/\hbar}
\label{qc}
\end{equation}
This is exactly of the second order WKB form (\ref{wkb2}), with $S_1$ giving
a prefactor that is a function of the combination $X=x^0\pm x^1$ ({\it
i.e.} gives a measure on the congruence of trajectories with fixed $P$),
and is in this sense localised around solutions with $X=\bar{X}$ as
expected.

The Wigner function for the quasiclassical state (\ref{qc}) can be easily
computed, and is equal to
\begin{eqnarray}
	F(x^\mu,p_\mu)&=&
	\int du \psi^*(x^\mu-u^\mu/2)
	\psi(x^\mu+u^\mu/2)e^{-ip_\mu u^\mu/\hbar}
	\cr
	&=& \left({4\hbar^3\pi\over\omega}\right)^{1/2}
	e^{-\omega (x^\pm -
	\bar{X}
	)^2/\hbar}
	e^{-({\bar{P}}-p_0)^2/\hbar\omega}
	\delta(p_0\pm p_1)
\end{eqnarray}
which is exactly of the form one expects for a left (right) moving
classical particle, localised in momentum and in $x^+$ $(x^-)$.

We could instead have worked directly in terms of eigenstates of the
operators $X$ and $P$. For a massless particle, the first order WKB states
are exact eigenstates of the operator $P$. A coherent superposition of such
states is clearly localised in both $P$ and $X$ and in this sense is
quasiclassical. It is easy to check (and follows from the superposition
principle) that a state of this kind has support only on $x^\mu$ within a
narrow tube around the trajectory defined by the mean values of $P$ and
$X$.

\bigskip

We are now ready to begin a rudimentary discussion of quasiclassical states
in quantum gravity.  In section 2, we shall review the standard
semiclassical approximation in quantum gravity, obtained by expanding the
Wheeler--DeWitt equation in powers of the Planck mass, and show how the WKB
approximation for the quantum gravitational degrees of freedom, and the
functional Schr\"odinger equation for the matter degrees of freedom
arise. In section 3, we shall discuss the application of the
Hamilton--Jacobi formalism to the gravitational field. Much of this
discussion is somewhat formal (at least in the context of an infinite
number of degrees of freedom), but it is useful for understanding first
order WKB states and their relation to gauge invariant quantization on the
space of all classical solutions. An appendix contains some details of how
the Hamilton--Jacobi formalism leads to the definition of gauge invariant
operators in a 1+1 dimensional cosmological model. In section 4, we run
through the three equivalent definitions of a quasiclassical state for the
gravitational field -- as a coherent state in terms of constants of the
motion, as a superposition of first order WKB states, and as a second order
WKB state with a localised prefactor -- making use of some of the ideas
discussed in section 3. It follows directly from the last definition that
the usual semiclassical approximation applies for this gravitational state
and yields the functional Schr\"odinger equation, but now effectively on a
single background spacetime. Section 5 contains a discussion of corrections
to the semiclassical approximation arising from an excessive sensitivity of
the matter evolution to very small changes in the background spacetime.

 \section{The semiclassical approximation}

Expanding the Wheeler--DeWitt and momentum constraint equations order by
order in the gravitational coupling constant $G$ leads to what is now well
known as the semiclassical approximation of quantum gravity. In this
section, we shall give a brief review of some of the large amount of work
on this subject (see \cite{lap,banks,jjh,vilenkin,kiefer,hartle,isham} and
references therein).

We shall ignore the details of the momentum constraint in the following
discussion, and assume that spatial diffeomorphism invariance is imposed at
all orders.  We shall assume a compact spatial topology, although this
discussion can be generalized to open spacetimes with well-defined
asymptotics (see for example the Appendix).

The Wheeler-DeWitt equation reads
\begin{equation}
	-16\pi G\hbar^2G_{ijkl}{\delta^2\Psi[f,h]
	\over\delta h_{ij}\delta h_{kl}}-{\sqrt{h}R\over
	16\pi G}\Psi[f,h]+{\cal H}_{matter}\Psi[f,h]=0
\label{a0}
\end{equation}
taking $c=1$. Consider expanding the state $\Psi$ as
\begin{equation}
\Psi=e^{i(S_0/G+S_1+GS_2+\cdots)/\hbar}
\label{a0a}
\end{equation}
where each $S_i$ is assumed to be of the same order. Eq. (\ref{a0}) can
then be expanded perturbatively in $G$.  The zeroth order equation (order
$1/G^{2}$) simply states that $S_0[f,h]=S_0[h]$ is independent of the
matter degrees of freedom. At the first non-trivial order (order $1/G$), we
find that
$S_0[h]$ must be a solution of the general relativistic Hamilton--Jacobi
equation \cite{gerlach,peres}
\begin{equation}
	{1\over 2}G_{ijkl}{\delta S_0\over
	\delta h_{ij}}{\delta S_0\over \delta h_{kl}} -2\sqrt{h}R=0.
\label{a1a}
\end{equation}
There is a large freedom in specifying solutions of the Hamilton--Jacobi
equation (\ref{a1a}), since it is necessary to specify a set of integration
constants. This is familiar from Hamilton--Jacobi theory, as we shall see
in the next section.

At the next order, we obtain an equation for $S_1[f,h]$.  It is convenient
to split $S_1[f,h]$ into two functionals $\chi[f,h]$ and $D[h]$, where
\begin{equation}
	{1\over D[h]}e^{iS_0[h]/\hbar G}
\label{a4}
\end{equation}
is the second order WKB approximation to the purely gravitational
Wheeler--DeWitt equation.  The equation for the WKB prefactor $D[h]$ is
\begin{equation}
	G_{ijkl}{\delta S_0[h]\over\delta h_{ij}}
	{\delta D[h]\over\delta h_{kl}}-
	{1\over 2}G_{ijkl}{\delta^2S_0[h]
	\over\delta h_{ij}\delta h_{kl}}D[h]=0.
\label{a1b}
\end{equation}
The remaining condition on $\chi[f,h]$,
\begin{equation}
	i\hbar G_{ijkl}{\delta S_0\over\delta h_{ij}}
	{\delta\chi[f,h]\over\delta h_{kl}}={\cal H}_{matter}\chi[f,h]
\label{a3}
\end{equation}
is an evolution equation for the functional $\chi[f,h]$ on the whole of
superspace (the space of all 3-geometries), and as such, its solution
requires initial data for $\chi[f,h]$ on a surface in superspace that is
crossed once by each classical trajectory defined by $S_0[h_{ij}]$.

Let us now recall how Eq. (\ref{a3}) is closely related to the functional
Schr\"odinger equation. Having specified the initial data, it can be solved
by the method of characteristics along the eikonal tracks on superspace
defined by $S_0$.  These tracks are the integral curves of the
Hamilton--Jacobi momenta, and are defined as solving the equations
\begin{equation}
\pi^{ij}=\delta S_0[h]/\delta h_{ij}\qquad{\rm or}\qquad {dh_{ij}({\bf
x},\tau) \over d\tau}=-2N({\bf x},\tau)K_{ij}({\bf
x},\tau)+\nabla_{(i}N_{j)}({\bf x},\tau)
\label{a4a}
\end{equation}
These are the set of solutions of Einstein's equations defined by the
Hamilton--Jacobi functional $S_0[h]$.  The solution of Eq.  (\ref{a4a})
requires a choice of integration constants and a choice of lapse and shift
functions $N({\bf x},\tau)$ and $N_i({\bf x},\tau)$.  The integration
constants specify different classical solutions while the lapse and shift
are just choices of coordinates on each of these spacetimes.  Along each
eikonal, Eq. (\ref{a3}) becomes the functional Schr\"odinger equation
\begin{equation}
i\hbar{\delta\chi[f,h]\over\delta\tau}=H_{matter}\chi[f,h]
\label{a3a}
\end{equation}
where $\tau$ is the time parameter corresponding to the chosen foliation.

As an aside, we remark that there are two potential integrability
conditions to worry about when solving Eq. (\ref{a3}) using the method of
characteristics. Firstly, Eq. (\ref{a4a}) can be integrated using different
lapse and shift functions, corresponding to using different coordinates on
the background spacetime. We expect (\ref{a3a}) to be covariant under
changes of coordinates, but this is not always the case. Integration with
different lapse and shift functions can lead to ambiguities in the
definition of $\chi[f,h]$, as has been discussed by various authors
\cite{kuchar,jackiw,nico}. We shall ignore this problem here.  Secondly,
there is the question of different integration constants in the solution of
Eq. (\ref{a4a}), corresponding to integration of (\ref{a3}) along different
classical spacetimes.  To the present order this causes no problems, since
in general there is at most one solution to Einstein's equations that
passes through any point in superspace and is compatible with a given
Hamilton-Jacobi functional $S_0[h]$ (that is the eikonals never
intersect). However, as we shall see in Sec. 5, evolution along
neighbouring eikonals is an important issue when considering corrections to
the semiclassical approximation.

It is possible to extend this approximate solution to the Wheeler--DeWitt
equation beyond second order. At that point the semiclassical picture of
field theory on a fixed background is lost. In principle, though, the
approximation scheme can be continued, separating the higher order WKB
approximations in the gravitational sector and corrections to the evolution
equation in the matter sector \cite{kiefer}. A general understanding of
these corrections is useful in determining the validity of the second order
approximation.

Although it might seem that obtaining the functional Schr\"odinger equation
is all there is to the semiclassical approximation, there is still the
question of whether the construction described above really describes
quantum field theory on a {\it single} spacetime.  Since each eikonal
trajectory defined by $S_0[h]$ is exactly classical, it does not make sense
to regard a WKB state as describing an ensemble of independent classical
solutions. Even if some form of decoherence is invoked, it is impossible
for a quantum state to describe an ensemble of strictly classical
spacetimes. To complete the semiclassical picture, we must understand the
nature of the gravitational background provided by the gravitational WKB
state.

\section{Hamilton--Jacobi formalism in quantum gravity}

Let us begin by considering the general relativistic Hamilton-Jacobi
equation (\ref{a1a}) (see Ref. \cite{komar}).  In order to specify a
solution of (\ref{a1a}), it is necessary to supply a series of constants of
integration which are usually called $\alpha$--parameters in
Hamilton--Jacobi theory (see for example Ref. \cite{goldstein}). Any
solution $S$ takes the form $S[h_{ij}(x);\alpha_I]$, where ${\alpha_I}$
represents an infinite number of integration constants (equivalent to two
field theory degrees of freedom in 3+1 dimensions \cite{isham}), and so
labels different solutions $S[h]$ of the Hamilton--Jacobi equation.  Given
a Hamilton--Jacobi functional $S[h;{\alpha_I}]$, the relation
\begin{equation}
	\pi^{ij}={\delta S[h_{ij};{\alpha_I}]\over\delta h_{ij}}
	\label{a5}
\end{equation}
gives the momenta conjugate to $h$ in terms of $h$ and
${\alpha_I}$. Eqs. (\ref{a5}), replacing $\pi^{ij}$ with $d{h}_{ij}/d\tau$
for some parameter $\tau$, are a set of first order differential equations
(c.f. Eqs. (\ref{a4a})), that yield a congruence of solutions to Einstein's
equations, but require a further set of integration constants to pick out a
particular solution. Equivalently, a classical solution can be fixed by
defining the values of a set of functionals
\begin{equation}
	{\beta_I}={\delta S[h_{ij};{\alpha_I}]\over\delta{\alpha_I}},
	\label{a5a}
\end{equation}
which are precisely the integration constants for (\ref{a5}), and then
solving for $h_{ij}$. The set of all $h_{ij}$ that satisfy this equation
form sheaf of trajectories in superspace defining a solution of Einstein's
equations (in the same way that (\ref{cons}) specified a set of coordinates
making up a classical trajectory). From either (\ref{a5}) or (\ref{a5a}) it
follows that a single solution of Einstein's equations requires a choice of
values for both the ${\alpha_I}$ and ${\beta_I}$ parameters. It also
follows that given a set of ${\alpha_I}$ parameters ({\it i.e.} a
Hamilton--Jacobi functional), a 3-geometry $h_{ij}$ fixes a unique value of
${\beta_I}$ for which the ${\alpha_I},{\beta_I}$ spacetime contains
$h_{ij}$.

Eqs. (\ref{a5}) and (\ref{a5a}) can be turned around to give a set of
conjugate {\it functionals} ${\alpha_I}[h_{ij},\pi^{ij}]$ and
${\beta_I}[h_{ij},\pi^{ij}]$ which are constants of the motion -- that is
they have weakly vanishing Poisson bracket with the Hamiltonian
constraint. These definitions provide a canonical transformation between
the $h_{ij}({\bf x})$ and $\pi^{ij}({\bf x})$ and the ${\alpha_I}$ and
${\beta_I}$, so that the Hamiltonian vanishes in the new coordinates. We
are free to write our theory in terms of these constants of the motion. The
${\alpha_I}$ and ${\beta_I}$ are coordinates and momenta on the physical
phase space\footnote{Of course the implicit equations (\ref{a5}) are
extremely difficult to solve in four dimensions, and so this discussion
should be regarded as somewhat formal in this sense.} (if we assume that we
have solved the momentum constraint) and so are the correct variables to
use for quantization according to the Dirac procedure.  They can be thought
of as parameterizing classical solutions of the Einstein equations
\cite{witten}, in the sense that fixing the values of
${\alpha_I}[h_{ij},\pi^{ij}]$ and ${\beta_I}[h_{ij},\pi^{ij}]$ yields a
classical solution simply by solving the equations
\begin{equation}
{\alpha_I}[h_{ij},\pi^{ij}]=\bar{\alpha_I},
\qquad{\beta_I}[h_{ij},\pi^{ij}]=\bar{\beta_I}.
\label{a6a} \end{equation}

In quantum gravity, classical correlations correspond precisely to the
specification of these constants of the motion. Of course it is
unreasonable to expect that all of the gravitational degrees of freedom
behave classically, but certainly a quantum state representing a classical
background spacetime must have a very small spread in those ${\alpha_I}$
and ${\beta_I}$ that are macroscopically observable. This simple
observation can be applied to great effect in understanding the
relationship between WKB states and quasiclassical states in quantum
gravity.

\section{Quasi--classical states in quantum gravity}

The standard Hamilton--Jacobi theory of the last section helps one to
understand first order WKB states of the form
$e^{iS[h_{ij};{\alpha_I}]}$. It is clear that a WKB state supplies the
values of the ${\alpha_I}$ parameters, since $S[h_{ij};{\alpha_I}]$ defines
a family of solutions to Einstein's equations with fixed ${\alpha_I}$ and
arbitrary ${\beta_I}$ (it is of course in this sense that the WKB state
contains information about a congruence of spacetimes).

Let us now imagine promoting the Poisson bracket algebra
\begin{equation}
\{\alpha_I,\beta_J\}\approx\delta_{IJ},\qquad\{{\cal H},{\alpha_I}\}\approx
\{{\cal H},{\beta_I}\}\approx 0
\label{a7}
\end{equation}
to an operator algebra in the space of functionals $\Psi[h_{ij}({\bf x})]$,
ignoring any anomalies or ordering ambiguities.  Although the Hamiltonian
vanishes in the ${\alpha_I}$ or ${\beta_I}$ representation, so that any
state $\Psi[{\alpha_I}]$ or $\Psi[{\beta_I}]$ is automatically a physical
state, this is a somewhat formal statement.  We can make more useful
observations by continuing to work in the metric representation.

The important thing to notice is that the first order WKB state
$e^{iS[h_{ij};\bar{\alpha_I}]/\hbar G}$ is an approximate eigenstate of the
operator $\hat{\alpha_I}[\hat h_{ij},\hat \pi^{ij}]$ with eigenvalue
$\bar{\alpha_I}$ in the sense that:
\begin{equation}
\hat{\alpha_I}e^{iS[h_{ij};\bar{\alpha_I}]/\hbar G}=
\bar{\alpha_I}e^{iS[h_{ij};\bar{\alpha_I}]/\hbar G} +O(\hbar G).
\label{a7a}
\end{equation}
Here we assume that some of the $\alpha_I$ are large compared to $\hbar G$,
which is equivalent to the assumption that underlies the semiclassical
approximation that the characteristic (length) scales of the gravitational
field are well above the Planck scale. It is easy to see why (\ref{a7a}) is
true under these conditions: We write ${\alpha_I}[h_{ij},\pi^{ij}]$ as an
operator by replacing the $\pi^{ij}$ by $i\hbar\delta/\delta h_{ij}$. Then
the leading order contribution to the rhs comes when all derivatives bring
down the exponent with its accompanying powers of $1/\hbar G$. In this
leading term, the derivatives are replaced by $\pi^{ij}=\delta
S[h_{ij},\bar{\alpha_I}]/\delta h_{ij}$ and by the definition of
$\alpha_I[h_{ij},\pi^{ij}]$, ${\alpha_I}[h_{ij},\delta
S[h_{ij},\bar{\alpha_I}]/\delta h_{ij}]=\bar{\alpha_I}$.  This is nothing
more than the standard first order WKB approximation in a different
guise. It is more difficult to compute the second order correction or
prefactor for an eigenstate of $\hat{{\alpha_I}}$. An eigenstate of
$\hat{\alpha_I}$ must have maximal uncertainty in ${\beta_I}$: As a
functional of $h_{ij}$, $\Psi_0[h_{ij},\bar{\alpha_I}]$ is damped only
where $h_{ij}$ is not found within any classical solution defined by
$\bar{\alpha_I}$ and ${\beta_I}$ for any ${\beta_I}$.

The above discussion shows clearly why a first order WKB state endowed with
a generic prefactor is not quasiclassical -- in general the constants of
the motion that define the classical correlations are not well
localised. The $\hat{\alpha_I}$ eigenstates provide an extreme example of
this. Another important aspect of the WKB approximation is evident -- that
it treats the conjugate variables $\alpha_I$ and $\beta_I$ asymmetrically,
so that the eikonal trajectories that lead to the functional Schr\"odinger
equation in the semiclassical approximation are defined by a single value
of the $\alpha_I$ but restricted in the $\beta_I$ only by the prefactor. On
the other hand, a classical spacetime is defined by a pair $\bar{\alpha_I}$
and $\bar{\beta_I}$ of gauge invariant quantities, and any classical
correlations imply a knowledge of both the ${\alpha_I}$'s and ${\beta_I}$'s
to a good degree of accuracy. It is clear that a quasiclassical quantum
state should be close to a coherent state, at least with respect to those
operators $\hat{\alpha}_I$ and $\hat{\beta}_I$
that correspond to macroscopic correlations.

Let us write exact eigenstates of $\hat{\alpha_I}$ which are also exact
physical states as $\Psi_{{\alpha_I}}[h_{ij}]$. A general physical
state is a superposition of eigenstates of $\hat{\alpha_I}$. A
quasiclassical state in quantum gravity is a coherent superposition of
$\hat{\alpha_I}$ eigenstates $\Psi_{\alpha_I}[h_{ij}]$:
\begin{equation}
\vert\Psi\rangle=\int d{\alpha_I}\; \omega({\alpha_I})
\vert{\alpha_I}\rangle \qquad{\rm or}\qquad \Psi[h_{ij}]=\int d{\alpha_I}\;
\omega({\alpha_I}) \Psi_{\alpha_I}[h_{ij}]
\label{sup}
\end{equation}
where $\omega({\alpha_I})$ is a distribution that ensures a close to
minimal uncertainty in both sets of observables ${\alpha_I}$ and
${\beta_I}$. Thus $\Psi[h_{ij}]$ in (\ref{sup})
has support only on a restricted region of superspace
centered around a classical solution, and is compatible with classical
correlations which effectively measure the gauge invariant quantities
${\alpha_I}$ and ${\beta_I}$.

Let us now consider how to write (\ref{sup}) in its second order WKB
approximation.  We write a state which approximates a classical spacetime
with parameters $\bar{\alpha_I}$ and $\bar{\beta_I}$ as
\begin{eqnarray}
\vert\Psi\rangle=\int d{\alpha_I}\;
e^{-i({\alpha_I}-\bar{\alpha_I})\bar{\beta_I}/\hbar G}
e^{-({\alpha_I}-\bar{\alpha_I})^2/2\hbar G}\vert{\alpha_I}\rangle
\label{111}
\end{eqnarray}
so that
$\omega({\alpha_I})=e^{-i({\alpha_I}-\bar{\alpha_I})\bar{\beta_I}/\hbar G}
e^{-(\alpha_I-\bar{\alpha_I})^2/2\hbar G}$. The phase in $\omega(\alpha_I)$
fixes the mean value of the $\beta_I$. Here we are working with $\alpha_I$
and ${\beta_I}$ normalized so that they have the same dimensions and that
$[\alpha_I,{\beta_J}]=i\hbar G\delta_{IJ}$. We assume that some $\alpha_I$
and ${\beta_I}$ are large compared to $\sqrt{\hbar G}$ so that there is a
large dimensionless parameter with respect to which we can perform the
expansion. This is related to the physical criterion that fluctuations
should be small compared to the characteristic scale of the solution. For
example, a cosmology with a maximal size of the order of the Planck scale
(see Sec. 6 for an example) should not be considered classical.  The choice
of $\omega(\alpha_I)$ in (\ref{111}) ensures that $\alpha_I$ and
${\beta_I}$ are localized to within $\sqrt{\hbar G}$ of their mean values
$\bar{\alpha_I}$ and $\bar{\beta_I}$.

In the metric representation we have by (\ref{a7a}) that
\begin{eqnarray}
\Psi_{\alpha_I}[h_{ij}] \cong e^{iS[h_{ij};\alpha_I]/\hbar G}
\end{eqnarray}
to first order, so that
\begin{equation}
\Psi_G[h_{ij}]\cong\int d{\alpha_I}\;
e^{-i(\alpha_I-\bar{\alpha_I})\bar{\beta_I}/\hbar G}
e^{-(\alpha_I-\bar{\alpha_I})^2/2\hbar G}e^{iS[h_{ij};\alpha_I]/\hbar G}.
\label{a8}
\end{equation}
The integration in (\ref{a8}) can be performed after expanding
$e^{iS[h_{ij};\alpha_I]}$ in powers of $\alpha_I-\bar{\alpha_I}$ (which is
forced to be small by the Gaussian). Keeping the relevant terms
contributing to $S_0$ and $D$ of Sec. 1, the result is
\begin{equation}
\Psi_G[h_{ij}]\cong e^{iS[\bar{\alpha_I}]/\hbar G}
e^{-(S'[\bar{\alpha_I}]-\bar{\beta_I})^2/2\hbar G}
\label{a9}
\end{equation}
where $S[\alpha_I]=S[h_{ij};\alpha_I]$, $S'[\alpha_I]=\delta
S[h_{ij};\alpha_I] /\delta{\alpha_I}$. To derive this result, we have used
the fact that the integration in (\ref{a8}) is over a Gaussian and that all
the eigenvalues of $\delta^{IJ}+i\delta^2 S/\delta\alpha_I\delta\alpha_J$
have positive real part. The first term in (\ref{a9}) is the first order
WKB approximation for $\alpha_I=\bar{\alpha_I}$, the center of the
Gaussian, and is the only rapidly oscillating term of order $1/\hbar G$.
The other term belongs in the next order correction, $S_1$ -- although it
may appear to be of the same order as the first term, the width of the
Gaussian makes it an order of magnitude lower. This second term damps
3-geometries $h_{ij}$ which are not compatible with
${\beta_I}=\bar{\beta_I}$. Although the $\alpha_I$ and ${\beta_I}$ damping
in this representation occur in different ways, the resulting state is
damped away from a narrow tube surrounding the mean spacetime given by
$\bar{\alpha_I}$ and $\bar{\beta_I}$. This is precisely what one expects
for a Gaussian superposition (\ref{sup}).

The definition of a semiclassical state is not limited to the case where
$\omega(\alpha_I)$ is an exact Gaussian. A general $\omega(\alpha_I)$ in
(\ref{sup}) will do equally well provided that it is peaked around some
$\bar{\alpha_I}$ and that its Fourier transform $\tilde{\omega}({\beta_I})$
is peaked around some $\bar{\beta_I}$ so that $\Delta \alpha_I,
\Delta{\beta_I} \sim \sqrt{\hbar G}$. Under these conditions one can write
(\ref{sup}) to a good approximation as
\begin{eqnarray}
\Psi[h_{ij}]=e^{iS[h_{ij},\bar{\alpha_I}]}\tilde{\omega}
\left(\frac{S'(\bar{\alpha_I})- \bar{\beta_I}}{\sqrt{\hbar G}}\right)
\end{eqnarray}
where $\tilde\omega$ contributes the damping in ${\beta_I}$, and both it
and its derivatives belong to $S_1$ or lower order terms.

If any quasiclassical superposition of first order WKB states is
approximately of the WKB form (\ref{a4}) once we take the prefactor into
account, then it fits into the expansion scheme described in Sec. 2.  The
matter portion of the WKB state, $\chi[f,h]$, is then still given by
solving the functional Schr\"odinger equation along characteristics, but
now these are restricted to be in the neighbourhood of the
$\bar{\alpha_I}$, $\bar{\beta_I}$ classical solution. The characteristics
are only those for $\bar{\alpha_I}$ and for all ${\beta_I}$'s within the
spread defined by the prefactor, so that the semiclassical approximation
still looks asymmetric with respect to $\alpha_I$ and ${\beta_I}$. However,
the difference between evolving $\chi[f,h]$ on any of the characteristics
generically belongs to lower order corrections because of the narrowness of
the tube. In this sense we can think only of solving the functional
Schr\"odinger equation on a mean spacetime defined by $\bar{\alpha_I}$ and
$\bar{\beta_I}$. Of course, in a more general sense, a coherent
superposition is symmetric in $\alpha_I$ and $\beta_I$, but this is hidden
in the asymmetry of the WKB approximation. Really, one should think of
solutions for any $\alpha_I$ and $\beta_I$ sufficiently close to the mean
values $\bar{\alpha_I}$ and $\bar{\beta_I}$ as being contained within a
coherent state, since the state has support on a tube around the mean
classical solution. The semiclassical approximation involves the choice of
one classical solution from within this tube as the classical background,
but there is no canonical choice of background from among all the classical
solutions contained within the tube defined by a particular quasiclassical
state (see Fig. 2).

\begin{figure}
\centerline{
\psfig{figure=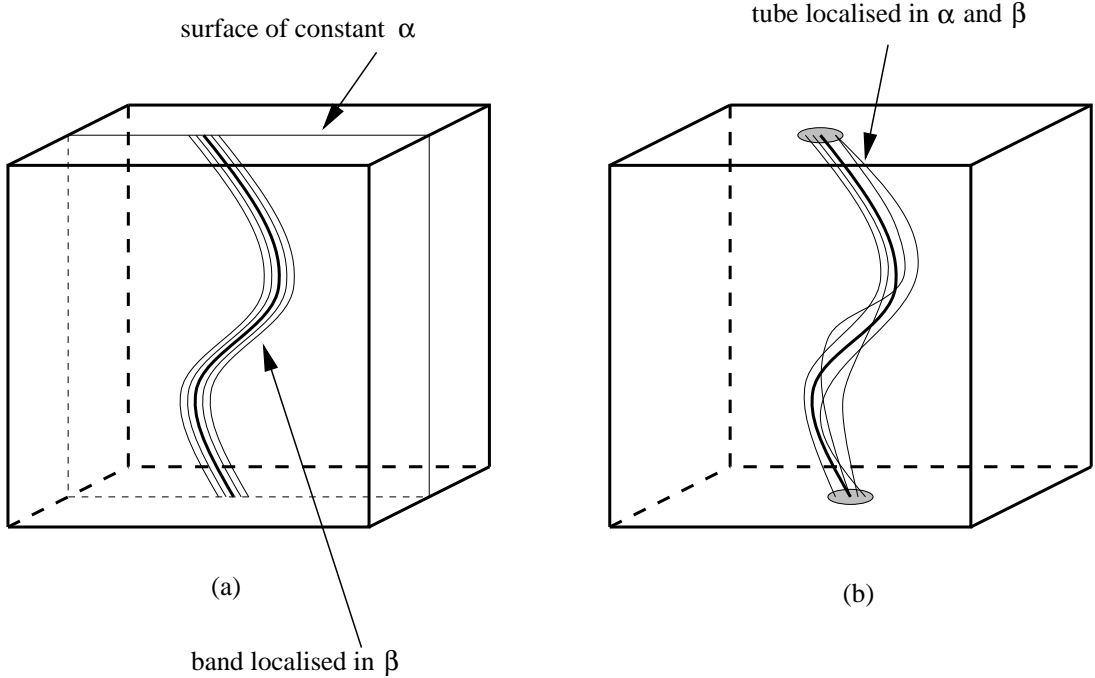,angle=-90,height=9cm}}
\begin{center}
\parbox{5in}{
\caption{\em For simplicity, in this figure, each classical trajectory is
represented by a single line, so that the box represents a gauge-fixed
version of superspace with a fixed choice of lapse and shift functions
defining each classical solution. The three dimensions of the box represent
the $\alpha_I$ and $\beta_I$ degrees of freedom and the time direction. (a)
shows a narrow band of trajectories with a fixed value of $\alpha_I$ and
with different values of $\beta_I$. The choice of $S_{HJ}$ fixes the value
of the $\alpha_I$ -- the surface shown -- and the prefactor then determines
the region of superspace on which the functional has support -- the band
within the surface. The bold line indicates a mean trajectory which is
surrounded by other sample trajectories. (b) more accurately represents a
quasi-classical state. Since we are considering a superposition of
$\alpha_I$ eigenstates, the wave functional has support off the surface of
constant $\alpha_I$, and throughout a narrow tube centered around the mean
trajectory. The lines represent classical trajectories with different
values of both $\alpha_I$ and $\beta_I$ in the neighbourhood of the
mean trajectory which is shown in bold.}}
\end{center}
\end{figure}

\section{Beyond the semiclassical approximation}

The fact that a quasiclassical wavefunctional for gravity (\ref{a9}) has
support on a tube in superspace rather than on a single eikonal track
provides some elementary intuition about going beyond the simple picture of
quantum field theory on a single background spacetime. As was mentioned at
the end of the last section, there is no preferred choice of classical
background from among all of those contained within the tube, and so if the
semiclassical approximation is to hold, the physics of the matter fields
had better be independent of this choice.

When one wants to talk about quantum gravity either outside or beyond the
semiclassical approximation, one has a fundamental problem: The lack of a
background spacetime, and of the notion of matter fields living on that
background makes life difficult, since it is unclear what is meant by a
unitary theory and how to define an inner product under these
circumstances.  Our discussion of the semiclassical approximation suggests
that some or all of these concepts make sense only to the same order as the
semiclassical approximation, but nonetheless form the basis of our current
description of nature.  The semiclassical approximation using
states is in this sense much the same as the sentiments expressed over the
years by Wheeler \cite{mtw}, since the Planck scale uncertainties in
$\alpha_I$ and ${\beta_I}$ can be related to fluctuations in the underlying
spacetimes which are generically on the Planck scale.

In this section we shall examine what we can learn qualitatively about
corrections arising from geometry fluctuations. A careful analysis of
corrections to the semiclassical approximation was carried out by Kiefer
and Singh \cite{ks} by expanding the semiclassical approximation to third
order. These authors derived a series of correction terms to the functional
Schr\"odinger equation, consisting of corrections in the integration of
matter fields along eikonal tracks and interference effects between eikonal
tracks. In Ref. \cite{ks} the former were shown to be small, but the
corrections due to interference, proportional to
\begin{equation}
{\delta\chi[f,h_{ij}]\over\delta h_{ij}}
\label{der}
\end{equation}
projected in a direction transverse to the eikonal trajectories, were
neglected. We shall take a geometrical approach to the computation of these
interference effects.

The basic quantity we wish to estimate is the state of matter on any given
hypersurface $\bar{h}_{ij}$
$$
\Psi_{\bar{h}_{ij}}[f] \equiv \Psi[f,\bar{h}_{ij}].
$$
To the order of the semiclassical approximation (equations (\ref{a1b}) and
(\ref{a3})), the evolution of this state is given by taking $\bar{h}_{ij}$
to be embedded only in the spacetimes labeled by $\bar{\alpha_I}$ and some
${\beta_I}$. Then one finds that $\Psi_{{h}_{ij}}[f]$ is a solution to
the functional Schr\"odinger equation on a fixed background.

A set of corrections to this approximation come from taking into account
the contributions from all the possible spacetimes labeled by $\alpha_I$
and ${\beta_I}$ which are not damped in the Gaussian state and which pass
through $\bar{h}_{ij}$.  A simple way to get qualitative information about
these corrections is to consider solving the functional Schr\"odinger
equation on all of these spacetimes\footnote{Not necessarily just those
eikonal tracks with $\alpha_I=\bar{\alpha_I}$ and $\beta_I$ within the
spread defined by the prefactor should be used, but all classical solutions
within the tube. Recall that although the WKB approximation picks out
eikonal tracks with a fixed value of $\alpha_I$ and different values of
$\beta_I$, the underlying state is approximately symmetric in $\alpha_I$
and $\beta_I$.} and comparing the properties of the solutions.  In order to
solve the functional Schr\"odinger equation for the set of spacetimes
defined by (\ref{a9}), it is necessary to give initial data on each of them
(that is on a surface in superspace transverse to the tube).  Sticking to
the philosophy that we wish to construct as classical a state as possible,
this initial data should be arranged to make the corrections to the
semiclassical approximation as small as possible.

If there are to be only Planck scale corrections to the semiclassical
approximation, the difference between Schr\"odinger evolution of matter
states on each of the spacetimes should be small, except at the Planck
scale.  If this difference were large, the results obtained to the order of
the semiclassical approximation would not be consistent.  It is clear that one
situation in which the semiclassical approximation breaks down unexpectedly is
when the evolution of the matter state has sensitive dependence on the
$\alpha_I$ and ${\beta_I}$ parameters.

When comparing Schr\"odinger evolutions on different backgrounds, we need
to compare different $\Psi[f,\bar{h}_{ij}]$ for the same
$\bar{h}_{ij}$. This must be done in a coordinate invariant fashion and
without reference to a spacetime. Some discussion of this issue can be
found in Ref. \cite{klmo}. The comparison of states on entire hypersurfaces
is fairly straightforward, but this does not provide a local description of
the interference effect. A local comparison of matter states can be made by
using matter correlation functions in the different spacetimes within the
Gaussian spread. Since the basic variables are 3-metrics, an insertion
point ({\it i.e.} an event) can only be defined geometrically by its
position on a 3-geometry (one might for example define a hypersurface by
its intrinsic geometry and then fix a point by its value of the intrinsic
curvature).  A correlation function should thus be defined by a 3-geometry
that contains all the insertion points, and by the location within the
given 3-geometry of the points. On a single classical background, the same
correlation function can be defined using many different choices of a
3-geometry that passes through all the insertion points. However, quantum
mechanically all of these choices are different. When we come to compute
the correlation function on different spacetimes within the tube, each
choice of 3-geometry identifies a different set of alternative classical
spacetimes in which the 3-geometry can be embedded, and so gives
different results (see Fig. 3(a)). Forcing an embedding of both surfaces
into a second spacetime is possible locally, but in this case the foliation
dependence occurs because the location of the insertion points in this
second spacetime depends on the choice of surface (see Fig. 3(b)). It
follows that the size of the corrections to the semiclassical approximation
obtained by comparing correlation functions depends on how one chooses to
foliate the original or mean spacetime. This dependence on foliation is
inevitable if we wish to compare local quantities.

\begin{figure}
\centerline{
\psfig{figure=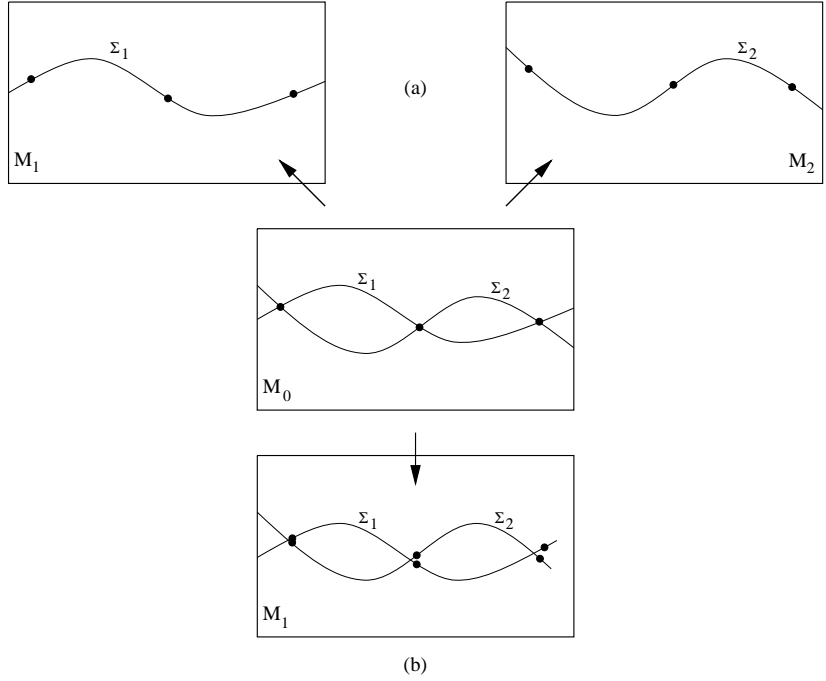,angle=-90,height=9cm}}
\begin{center}
\parbox{5in}{
\caption{\em A pair of surfaces $\Sigma_1$ and $\Sigma_2$ in a classical
spacetime $M_0$ are each used to define three insertion points. The
computation of a matter state at the points is independent of foliation on
$M_0$. (a) When quantum gravitational corrections are computed, each
surface must be embedded in a different spacetime, since two 3-geometries
determine a unique solution of Einstein's equations. Computing a
correlation function after evolution in $M_1$ to $\Sigma_1$ or in $M_2$ to
$\Sigma_2$ will in general give different answers. (b) If one forcibly
embeds both $\Sigma_1$ and $\Sigma_2$ in $M_1$ locally, without worrying
about global fitting, then the points will not coincide when defined with
respect to the two different surfaces. }}
\end{center}
\end{figure}

However small the dependence on the choice of foliation, for corrections to
the functional Schr\"odinger equation, coordinate invariance is
lost\footnote{This foliation dependence is independent of the anomalies
discussed in Sec. 2 and in Refs. \cite{kuchar,jackiw,nico}.}. In some
dramatic situations where evolution with respect to certain foliations is
very sensitive to small differences in $\alpha_I$ and $\beta_I$, this can
lead to very different conclusions about the size of quantum gravity
effects in different foliations (or frames of reference). This might look
puzzling since we started off with the Wheeler--DeWitt equation which is
supposed to impose coordinate invariance.  Recall, however, that the
familiar notion of coordinate invariance comes from the covariance of
matter evolution on a fixed background spacetime, which is valid only in
the semiclassical approximation.  To this order, observations are
independent of a choice of foliations of the mean background spacetime.  It
is this notion which breaks down when one takes into account the geometry
fluctuations which are higher order corrections. This is because the
meaning of the Wheeler--DeWitt equation is different at this next order,
since the notion of diffeomorphism invariance is now a property of the
combined matter--gravity system, not just of matter on a fixed background.

Generally one might expect interference effects of the kind we have
described to always be restricted to the Planck scale in any reasonable
models. However, it is important to note that we cannot trust our usual
intuition about quantum gravity when quantifying these effects, since they
are not defined by the quantities one normally associates with a breakdown
in the semiclassical approximation such as large curvature or
backreaction. What makes this discussion particularly relevant is that in
black hole physics, quantum gravity effects that normally occur at the
Planck scale can be magnified to the classical scale by the apparently
chaotic behaviour of functional Schr\"odinger evolution on certain
hypersurfaces close to the horizon.  In a variety of models, it has been
shown that the evolution of matter states on foliations corresponding to
outside observers close to a black hole horizon is extremely sensitive to
very small fluctuations in the background geometry, so that the projection
of (\ref{der}) transverse to the classical solution is very large (see
\cite{klmo}, and also \cite{verlindes} for a related discussion). Since
large corrections are present in this case for only a very specific
foliation, this indicates that a breakdown in coordinate invariance
accompanies the breakdown in the semiclassical approximation. In an
effective description, the results of certain sets of observations near the
black hole horizon are not covariant, providing an explanation for the
quantum gravitational origin of the spacetime complementarity proposed by
't Hooft \cite{thooft} and Susskind \cite{susskind}.

 \section{Conclusions}

We have shown that the appropriate state to consider as quasiclassical in
quantum gravity is a superposition of WKB type states which is peaked
around some values of the reduced phase space variables, with close to
minimal uncertainty in the reduced phase space variables. A first order WKB
state on the other hand is an approximate eigenstate of half of these
variables and hence not adequate.  When matter is present, this is
perfectly compatible with the derivation of the Schr\"odinger equation from
the Wheeler--DeWitt constraint. This makes more precise the meaning of a
semiclassical approximation in quantum gravity.

Using a superposition of WKB states, we were able to give a heuristic
treatment of higher order effects due to the quantum nature of the
background geometry. This allowed us to see that the semiclassical
approximation can be inconsistent because of the sensitivity of matter
propagation to small fluctuations in the background geometry. An example of
this type of situation is given by the breakdown of a semiclassical
description of matter propagating on a black hole spacetime
\cite{samir,klmo}.  We also showed that as a consequence of quantum
fluctuations, coordinate invariance is lost on the Planck scale, and in
certain cases, such as near the black hole, this extends to macroscopic
scales.

We have not discussed in this paper how it is that a system described by
the Wheeler--DeWitt equation comes to find itself in the particular state
that exhibits semiclassical behaviour. There is in principle no kinematical
reason to prefer this state over any other.  Perhaps the most likely answer
to this question is that decoherence effectively forces any state to a
configuration in which observations are equivalent to those within the
Gaussian state. However, the important point is that decoherence cannot
drive a state towards any configuration for which the background spacetime
is more classical than the one we have described.

\section*{Acknowledgments}

We are grateful to S. Carroll, D. Giulini, J. Halliwell, R. Jackiw,
E. Keski--Vakkuri, J. Samuel, T. P. Singh and A. Vilenkin for helpful
discussions. After this work was completed, the authors learned that
results overlapping with the discussion in the appendix have been obtained
independently by J-G. Demers \cite{demers}.

\section*{Appendix: A two-dimensional example}

The use of Hamilton--Jacobi theory to reduce to the physical degrees of
freedom was discussed rather abstractly in Sec. 3. It is instructive to
illustrate this using a general 1+1 dimensional dilaton gravity model. The
model we shall consider was discussed by Louis-Martinez {\it et al}
\cite{LGK}, and we shall make extensive use of their results. Related work
on open and closed spacetimes can be found in Ref. \cite{1+1}.

\subsection*{A.1. Classical theory: closed universe}

Let us focus primarily on the case of a closed cosmology.  Consider the
action,
\begin{eqnarray}
S=\int_{\cal M} d^2x \sqrt{-g} \left[\phi R-V(\phi)\right]
\end{eqnarray}
where $x$ is a periodic coordinate (with period $2\pi$), so that ${\cal
M}=S^1\times R$. This model reduces to minisuperspace and string inspired
models for particular choices of $V(\phi)$. For example, a constant
potential $V(\phi)=-4\lambda^2$ gives the closed universe version
\cite{russo} of the CGHS model \cite{cghs}.

Working with the parameterization
\begin{eqnarray}
ds^2=e^{2\rho}\left(-N^2 dt^2+(dx+N_\perp dt)^2\right)
\end{eqnarray}
for the metric $g_{\mu\nu}$, the canonical variables are $\rho(x)$ and
$\phi(x)$, with conjugate momenta
\begin{eqnarray}
&&\Pi_\phi={2\over N}\left(N_\perp\rho'+N_\perp'-\dot{\rho} \right)
\label{5} \\ &&\Pi_\rho=-{2\over N}\left(\dot{\phi}-N_\perp\phi'\right)
\label{6}
\end{eqnarray}
while the lapse and shift functions $N$ and $N_\perp$ are Lagrange
multipliers. As usual, the Hamiltonian is just a sum of the Hamiltonian and
diffeomorphism constraints
\begin{eqnarray}
H=\int dx\left[N{\cal H}+N_\perp{\cal H}_\perp\right]
\end{eqnarray}
where
\begin{eqnarray}
&& {\cal H}=2\phi''-2\phi'\rho'-{1\over
2}\Pi_\rho\Pi_\phi+e^{2\rho}V(\phi),
\label{h1} \\ && {\cal H}_\perp=\rho'\Pi_\rho+\phi'\Pi_\phi-\Pi_\rho'.
\label{h2} \end{eqnarray}

The Hamilton-Jacobi equation reads
\begin{eqnarray}
g[\phi,\rho]+{\delta S\over\delta\phi} {\delta S\over\delta\rho}=0
\end{eqnarray}
where
\begin{eqnarray}
g[\phi,\rho]=-4\phi''+4\phi'\rho'-2e^{2\rho} V(\phi).
\end{eqnarray}
This is solved by the functional
\begin{equation}
S[\phi,\rho;C]=2\int dx \left\{Q_C+\phi'\ln\left[{2\phi'-Q_C\over
2\phi'+Q_C} \right]\right\}
\label{b1}
\end{equation}
where $C$ is a constant,
\begin{eqnarray}
Q_C[\phi,\rho]=2\left[(\phi')^2+\left(C+j(\phi)\right)e^{2\rho}
\right]^{1\over 2}
\end{eqnarray}
and
\begin{eqnarray}
{dj(\phi)\over d\phi}=V(\phi).
\end{eqnarray}
(\ref{b1}) is also invariant under spatial diffeomorphisms.

In the Hamilton--Jacobi functional (\ref{b1}), we see the presence of a
parameter $C$ which is the $\alpha$ parameter in this problem. To deduce
$C$ as a functional of $\phi$, $\rho$ and their conjugate momenta, we must
invert the relations
\begin{eqnarray}
{\delta S\over\delta\phi}={g[\phi,\rho]\over Q_C[\phi,\rho]},\qquad {\delta
S\over\delta\rho}=Q_C[\phi,\rho].
\end{eqnarray}
These lead to the definition
\begin{equation}
C=e^{-2\rho}\left({1\over 4}\Pi_\rho^2-(\phi')^2\right) -j(\phi)
\label{cee}
\end{equation}
Similarly we can define the quantity $\beta$, which we shall call $P$
following \cite{LGK}, as
\begin{equation}
P={\delta S\over\delta C}=- \int dx {2e^{2\rho}\Pi_\rho\over
\Pi_\rho^2-4(\phi')^2}.
\label{pee}
\end{equation}
It is easy to check that $C$ and $P$ are conjugate and that they have
weakly vanishing Poisson brackets with the constraints.

{}From Eq. (\ref{b1}) for the Hamilton--Jacobi functional, we can solve the
classical equations of motion, using the relations
\begin{eqnarray}
\Pi_\phi={\delta S\over\delta\phi},\qquad\Pi_\rho= {\delta S\over\delta\rho}
\end{eqnarray}
and Eqs. (\ref{5}) and (\ref{6}).

For a constant potential $V(\phi)=-4\lambda^2$, and taking $\sigma=1$ and
$M=0$, there are homogeneous solutions
\begin{eqnarray}
ds^2={\lambda^2 P^2\over \pi^2}e^{-2\lambda^2Pt/\pi}(-dt^2+dx^2)
\end{eqnarray}
and
\begin{eqnarray}
\phi={C\over 4\lambda^2}-e^{-2\lambda^2Pt/\pi}
\end{eqnarray}
for all values of $C$ and $P$.

Armed with a solution of the Hamilton--Jacobi equation, it is interesting
	to look briefly at the quantisation of this model. We may promote
	${\Pi}_\phi$ and $\Pi_\rho$ to operators \begin{equation}
	\hat{\Pi}_\phi(x)=-i\hbar{\delta\over\delta\phi(x)},\qquad
	\hat{\Pi}_\phi(x)=-i\hbar{\delta\over\delta\phi(x)}
\end{equation}
With any ordering of the constraints, the first order WKB approximation is
	of course given by \begin{equation}
	\Psi[\phi,\rho]=e^{iS[\phi,\rho;C]/\hbar}
\label{fowkb}
\end{equation}
There is however a convenient choice of operator ordering \cite{LGK}
\begin{eqnarray}
	{\cal H} &=& {1\over 2}g[\phi,\rho] + {1\over 4}Q_{C_0}[\phi,\rho]
	\hat{\Pi}_\phi Q_{C_0}^{-1}[\phi,\rho]\hat{\Pi}_\rho, \cr
	{\cal H}_\perp &=& \phi' \hat{\Pi}_\phi +\rho'\hat{\Pi}_\rho
	-\hat{\Pi}_\rho'
\end{eqnarray}
which depends on a parameter $C_0$. With this choice of ordering
(\ref{fowkb}) is an exact solution of the constraint equations in the
particular case where the value of $C$ in $S[\phi,\rho;C]$ is equal to
$C_0$.

Let us now use this exact solution as a first order WKB solution and
	continue the WKB expansion (as we did for the relativistic particle
	where {\it all} the first order WKB states $e^{iPx^\pm/\hbar}$ were
	exact solutions of the Klein--Gordon equation). Writing
	\begin{equation} \Psi[\phi,\rho]=e^{iS[\phi,\rho;C_0]/\hbar
	+iS_1[\phi,\rho]}
\end{equation}
the equation for $S_1$ is \begin{equation}
	{\delta S_1[\phi,\rho]\over \delta \phi}=-{g[\phi,\rho]\over
	Q_{C_0}^2[\phi,\rho]}{\delta S_1[\phi,\rho]\over \delta\rho}
	\label{sowkb}
\end{equation}
Note that this equation is not solved by $S[\phi,\rho;C]$ because of the
minus sign. We can however find an exact solution of this equation by
remembering that we expect the prefactor $-i\ln S_1$ to be a weight
function over different values of $P$, the conjugate to $C$. Indeed, if we
define the functional
\begin{equation}
P[\phi,\rho]\equiv -{1\over 2}\int dx {Q_{C_0}[\phi,\rho]\over (C+j(\phi))}
\end{equation}
where we have replaced $\Pi_\rho$ in (\ref{pee}) by its classical value
given by the Hamilton--Jacobi functional, then an arbitrary function of $P$
solves (\ref{sowkb}) and the WKB approximation centered around $C_0$ is of
the form
\begin{equation}
	\Psi[\phi,\rho]={1\over D[P[\phi,\rho]]}e^{iS[\phi,\rho;C=C_0]}
\label{soapp}
\end{equation}
If we take $D[P[\phi,\rho]]$ to be a Gaussian, then we obtain an explicit
expression for the quasiclassical state to this order. Note that for a
general $D[P[\phi,\rho]]$, (\ref{soapp}) is no longer an exact solution of
the constraint equation, meaning that at least some of the $S_i$, $i\ge 2$
must be non-zero.

\subsection*{A.2. Classical theory: spacetimes with boundary}

The case of an open universe has been studied by various authors
\cite{1+1}.  It has been shown that the variable $C$ is related to the ADM
mass of the spacetime, while $P$, integrated throughout a spacelike slice,
is related to the time at infinity (or more precisely, to the
synchronization between times at infinity).  These results are in keeping
with the much earlier work of Regge and Teitelboim \cite{rt} on conserved
charges in canonical quantum gravity in open universes.

It is interesting to note that while $P$ is associated with a constant of
the motion as described in Refs. \cite{1+1}, a closely related quantity
provides a local geometric definition of time for different hypersurfaces
within a static spacetime.  Consider any 1-geometry associated with a
hypersurface in a static classical solution.  It can be intrinsically
described by the function $\phi(s)$, where $s$ is the proper distance along
the hypersurface measured from some base point $B$ at infinity, and $\phi$
is the value of the dilaton field. Let $\phi_0(s)$ be the function defining
a constant time surface $t=t_1$ passing through $B=\phi_0(0)$. Consider now
a set of hypersurfaces passing through $B$ that are defined by $\phi_i(s)$
which differ from $\phi_0(s)$ only in some finite interval $0<s<s_0$. For
$s>s_0$, $\phi_i(s)$ also define constant time hypersurfaces but at some
time $t_i=t_0+\Delta t_i$.

Using (\ref{cee}) to give $\Pi_\rho$ in terms of $C$ and $\phi$, we can
define a quantity
\[
T_i(S)=-\int_{0}^{S} ds {\left[\left({d\phi_i\over
ds}\right)^2+C+j(\phi_i)\right]^{1/2} \over (C+j(\phi_i))}
\]
closely related to $P$. Here $S>s_0$ so that the integration extends well
into the region where $\phi_i(s)=\phi_0(s)$.

Now in a static coordinate system
\[
ds^2=e^{2\rho}(-dt^2+dx^2)
\]
we can compute the change in the time coordinate along any hypersurface
using
\[
\left({dt\over dx}\right)^2=1-{e^{-2\rho} \left({d\phi_i\over dx}\right)^2
\over \left({d\phi_i\over ds}\right)^2}.
\]
Since in the static coordinate system $\Pi_\rho=0$, it follows that
\[
\left({dt\over dx}\right)^2=1+{C+j(\phi_i)\over \left({d\phi_i\over
ds}\right)^2}.
\]
{}From this expression we deduce that
\[
\Delta t_i=\int dx {\left[\left({d\phi_i\over ds}\right)^2
+C+j(\phi_i)\right]^{1/2}\over \left({d\phi_i\over ds}\right)} = \int
^S_{0} ds {\left[\left({d\phi_i\over ds}
\right)^2+C+j(\phi_i)\right]^{1/2}\over e^{\rho}\left({d\phi_i\over
ds}\right)}.
\]
By definition $\Delta t$ is zero for $\phi_0$, but is non-zero for any
$\phi_i(s)$ over the region $0<s<s_0$.

The connection between $\Delta t_i$ and $T_i(S)$ emerges by noting that any
static solution has
\[
e^{\rho}=c_0 \left({d\phi\over ds}\right)
\]
where $c_0$ is some constant of proportionality (this can be shown using
Eqs. (\ref{h1}) and (\ref{cee})). From this it follows that
\[
\Delta t_i(S)={T_i(S)\over c_0}.
\]

\end{document}